\documentclass[reprint, aps, prd, showpacs, superscriptaddress, floatfix]{revtex4-2}
\usepackage{graphicx}

\usepackage{amsmath}

 \usepackage{orcidlink} 
 
\begin{document}

\title{Quantum-Corrected Entropy Bounds on Black Hole Merger Efficiency}
\author{Jiswin Varghese}
\email{jiswinvarghese@gmail.com}
\affiliation{Department of Physics, Government College, Kattappana, Idukki, Kerala 685508, India}
\orcidlink{0009-0008-7185-8983} 
\begin{abstract}
We examine the entropy balance of black hole mergers in the presence
of quantum gravity corrections described by the Generalized Uncertainty
Principle (GUP). Focusing on the coalescence of two non-spinning
Schwarzschild black holes with unequal masses, we analyze the
thermodynamic evolution from the initial binary system to the final
merger remnant, explicitly accounting for energy loss through
gravitational-wave emission. Using a logarithmically corrected entropy
motivated by a quadratic GUP, we derive a generalized entropy change as
a function of the gravitational-wave efficiency and the mass ratio of
the merging components. The classical area theorem is recovered in the
appropriate limit, while the GUP correction introduces a quantum-modified
entropy bound that constrains the maximum allowed gravitational-wave
energy emission. This bound is most restrictive for nearly equal-mass
mergers, indicating an enhanced sensitivity of symmetric systems to
quantum gravity effects.
\end{abstract}


\maketitle
\section{Introduction}
The coalescence of compact objects represents one of the most energetic phenomena in the Universe. In particular, black hole mergers provide a unique arena to test the dynamical and thermodynamical properties of strong-field gravity. The direct detection of gravitational waves from binary black hole mergers by the LIGO–Virgo collaboration has firmly established these systems as astrophysical realities and has opened new opportunities to probe fundamental aspects of gravitation in highly nonlinear regimes \citep{Abbott2016,Abbott2019}.

From a theoretical perspective, black hole mergers are constrained by classical results in general relativity, most notably Hawking’s area theorem, which states that the total event horizon area of a system of black holes cannot decrease in any classical process \citep{Hawking1971}. When interpreted thermodynamically, this theorem is equivalent to the statement that the total entropy of the system, identified with the Bekenstein–Hawking entropy, must increase during a merger \citep{Bekenstein1973}. Energy carried away by gravitational waves reduces the final mass of the remnant black hole, yet the area theorem ensures that the entropy of the final state exceeds that of the initial binary configuration.

Despite its robustness, the area theorem is derived within the framework of classical general relativity and does not account for possible quantum gravity effects. Various approaches to quantum gravity suggest that the Bekenstein–Hawking entropy receives corrections, often logarithmic in the horizon area, arising from quantum or Planck-scale physics \citep{Kaul2000}. Among phenomenological frameworks incorporating such effects, the Generalized Uncertainty Principle (GUP) has received considerable attention. Motivated by Gedanken experiments in quantum gravity and string theory, the GUP introduces a minimal length scale and leads naturally to corrections in black hole thermodynamics \citep{Adler2001,Scardigli1999}.

The implications of GUP-modified entropy have been explored in a variety of contexts, including black hole evaporation, remnant formation, and horizon thermodynamics \citep{Nozari2008,Ali2012}. However, comparatively less attention has been paid to the role of such corrections in dynamical processes involving multiple black holes, particularly mergers accompanied by gravitational-wave emission. Given the precision with which the energy radiated in gravitational waves can now be estimated from observations and numerical relativity, it is natural to ask whether quantum-corrected entropy bounds impose additional constraints on merger processes.

In this work, we investigate the entropy balance of black hole mergers in the presence of GUP-induced corrections. We consider the coalescence of two non-spinning Schwarzschild black holes with unequal masses and explicitly account for energy loss through gravitational radiation. By employing a logarithmically corrected entropy motivated by a quadratic GUP, we derive a generalized entropy change for the merger process and analyze the conditions under which the generalized second law is satisfied. Our approach is fully analytical and does not rely on detailed numerical simulations or observational data fitting. The results reveal that quantum gravity corrections introduce a modified entropy bound on the gravitational-wave efficiency, with the strongest constraints arising for nearly equal-mass mergers. This suggests that black hole merger thermodynamics provides a simple and consistent framework for exploring potential quantum corrections to classical gravitational dynamics.

\section{Classical Entropy Bound and GW Efficiency}
We consider the merger of two non-spinning Schwarzschild black holes with initial masses $M_1$ and $M_2$, forming a single remnant black hole after the emission of gravitational radiation. The total energy radiated in gravitational waves is parametrized by the dimensionless efficiency $\epsilon$, such that the final mass of the remnant is
\begin{equation}
M_f = (1 - \epsilon)(M_1 + M_2).
\end{equation}

In classical general relativity, the entropy of a Schwarzschild black hole is proportional to the area of its event horizon,
\begin{equation}
S_{\mathrm{BH}} = \frac{A}{4} = 4\pi M^2,
\end{equation}
where natural units $G = c = \hbar = k_B = 1$ are used throughout this work.

The total initial entropy of the binary system is therefore
\begin{equation}
S_i = 4\pi \left(M_1^2 + M_2^2\right),
\end{equation}
while the entropy of the final remnant black hole is
\begin{equation}
S_f = 4\pi M_f^2 = 4\pi (1 - \epsilon)^2 (M_1 + M_2)^2.
\end{equation}

The classical generalized second law requires that the total entropy does not decrease during the merger process,
\begin{equation}
\Delta S_{\mathrm{cl}} \equiv S_f - S_i \geq 0.
\end{equation}

To express this condition in a dimensionless and scale-independent form, we introduce the mass ratio
\begin{equation}
q \equiv \frac{M_2}{M_1}, \qquad 0 < q \leq 1,
\end{equation}
and factor out the overall mass scale $M_1^2$. The normalized entropy change becomes
\begin{equation}
\frac{\Delta S_{\mathrm{cl}}}{4\pi M_1^2}
= (1 - \epsilon)^2 (1 + q)^2 - (1 + q^2).
\end{equation}

The condition $\Delta S_{\mathrm{cl}} \geq 0$ yields an upper bound on the gravitational-wave efficiency,
\begin{equation}
\epsilon \leq \epsilon_{\max}^{\mathrm{cl}}(q),
\end{equation}
where the maximum allowed efficiency is obtained by solving
\begin{equation}
(1 - \epsilon_{\max}^{\mathrm{cl}})^2 (1 + q)^2 = (1 + q^2).
\end{equation}

This classical bound depends solely on the mass ratio of the binary and is independent of the overall mass scale. In the equal-mass limit $q = 1$, the bound reduces to
\begin{equation}
\epsilon_{\max}^{\mathrm{cl}} = 1 - \frac{1}{\sqrt{2}},
\end{equation}
while for highly asymmetric mergers $q \ll 1$, the constraint becomes progressively weaker. The classical entropy bound thus provides a natural thermodynamic limit on the efficiency of gravitational-wave emission during black hole mergers.

This classical framework serves as the reference against which quantum-gravity--induced corrections will be evaluated in the following section.

\section{GUP-Corrected Entropy and Dimensionless Formulation}

Several approaches to quantum gravity suggest the existence of a minimal measurable length, leading to a modification of the standard Heisenberg uncertainty principle \cite{Adler2001}. A common phenomenological realization is the quadratic GUP:
\begin{equation}
\Delta x \Delta p \geq \frac{1}{2} \left[ 1 + \alpha (\Delta p)^2 \right],
\end{equation}

where $\alpha$ encodes quantum-gravity effects. By identifying the uncertainty in position $\Delta x$ with the horizon radius $r_s = 2M$, we solve for the momentum uncertainty to find the GUP-modified Hawking temperature:
\begin{equation}
T_{GUP} \approx \frac{1}{8\pi M} \left[ 1 + \frac{\alpha}{16M^2} \right]
\end{equation}
Using the first law of black hole mechanics, $dS = \frac{dM}{T}$, the entropy is obtained by integration:
\begin{equation}
S_{GUP} = \int 8\pi M \left( 1 + \frac{\alpha}{16M^2} \right)^{-1} dM \approx 4\pi M^2 - \frac{\alpha}{2} \ln(M^2) + C
\end{equation}
By expressing the result in terms of the horizon area $A = 16\pi M^2$ and absorbing constants into the parameter $\alpha$ (consistent with \cite{Kaul2000, Medved2004, Majumder2011}), we arrive at the corrected entropy:

\begin{equation}
S_{\mathrm{GUP}} = 4\pi M^2 + \alpha \ln \left( 4\pi M^2 \right),
\end{equation}
where $\alpha$ is treated as a dimensionless parameter. In the limit $\alpha \rightarrow 0$, the classical Bekenstein-Hawking entropy is recovered \cite{Bekenstein1973}.

We now consider the entropy balance for a binary black hole merger within this GUP-modified framework. The initial entropy of the system is given by
\begin{equation}
S_i^{\mathrm{GUP}} = 4\pi \left( M_1^2 + M_2^2 \right)
+ \alpha \ln \left( 16\pi^2 M_1^2 M_2^2 \right),
\end{equation}
while the entropy of the final remnant black hole is
\begin{equation}
S_f^{\mathrm{GUP}} = 4\pi (1 - \epsilon)^2 (M_1 + M_2)^2
+ \alpha \ln \left[ 4\pi (1 - \epsilon)^2 (M_1 + M_2)^2 \right].
\end{equation}

The total GUP-corrected entropy change associated with the merger process is therefore
\begin{equation}
\Delta S_{\mathrm{GUP}} \equiv S_f^{\mathrm{GUP}} - S_i^{\mathrm{GUP}}.
\end{equation}

To facilitate a scale-independent analysis, we again introduce the mass ratio
\begin{equation}
q \equiv \frac{M_2}{M_1},
\end{equation}
and normalize the entropy change by factoring out $4\pi M_1^2$. After simplification, the dimensionless GUP-corrected entropy change can be written as
\begin{equation}
\frac{\Delta S_{\mathrm{GUP}}}{4\pi M_1^2}
= (1 - \epsilon)^2 (1 + q)^2 - (1 + q^2)
+ \frac{\alpha}{4\pi M_1^2}
\ln \left[
\frac{(1 - \epsilon)^2 (1 + q)^2}{q^2}
\right].
\end{equation}

Since the overall mass scale appears only as a multiplicative factor in the logarithmic term, it is convenient to absorb all dimensionful constants into an effective dimensionless GUP parameter. We therefore define
\begin{equation}
\tilde{\alpha} \equiv \frac{\alpha}{4\pi M_1^2},
\end{equation}
which allows the entropy change to be expressed in a fully dimensionless form,
\begin{equation}
\Delta \tilde{S}_{\mathrm{GUP}}
= (1 - \epsilon)^2 (1 + q)^2 - (1 + q^2)
+ \tilde{\alpha}
\ln \left[
\frac{(1 - \epsilon)^2 (1 + q)^2}{q^2}
\right].
\end{equation}

The generalized second law requires $\Delta \tilde{S}_{\mathrm{GUP}} \geq 0$, which now yields a quantum-modified upper bound on the gravitational-wave efficiency,
\begin{equation}
\epsilon \leq \epsilon_{\max}^{\mathrm{GUP}}(q, \tilde{\alpha}).
\end{equation}

In the limit $\tilde{\alpha} \rightarrow 0$, the classical entropy bound derived in Section~2 is recovered. For finite positive values of $\tilde{\alpha}$, the allowed parameter space for gravitational-wave emission is modified, leading to potentially observable deviations from classical predictions. These effects are explored quantitatively in the following section.

\section{Results and Discussion}
We have analyzed the entropy balance of binary black hole mergers by incorporating gravitational-wave energy loss and quantum corrections motivated by the Generalized Uncertainty Principle. The results are presented in Figures~1--3 and are discussed below.
\begin{figure}
\centering
\includegraphics[width=0.9\columnwidth]{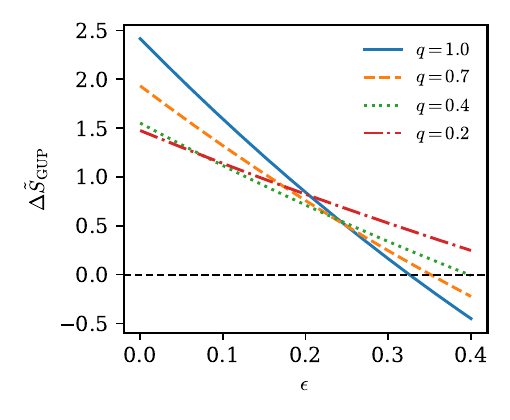}
\caption{
Dimensionless entropy change $\Delta\tilde S$ as a function of the gravitational-wave efficiency $\epsilon$ for several mass ratios $q = M_2/M_1$. Solid curves correspond to the GUP-corrected entropy with $\tilde{\alpha} = 0.3$. The entropy decreases monotonically with increasing $\epsilon$, and the slope becomes steeper for more asymmetric mergers. The horizontal line $\Delta\tilde S = 0$ indicates the thermodynamic censorship boundary separating allowed and forbidden merger configurations.
}
\end{figure}

Figure~1 shows the dimensionless entropy change as a function of the gravitational-wave efficiency for several mass ratios. For all configurations, the entropy decreases monotonically with increasing $\epsilon$, reflecting the reduction of the final horizon area due to radiated energy. The slope of this decrease becomes steeper for larger mass asymmetry, indicating that mergers with a dominant primary black hole possess a smaller entropy budget and are therefore more sensitive to gravitational-wave losses. In all cases, the entropy initially remains positive, ensuring consistency with the generalized second law, until a critical efficiency is reached. The horizontal line $\Delta\tilde S = 0$ marks the thermodynamic censorship boundary beyond which the merger configuration becomes forbidden.

\begin{figure}
\centering
\includegraphics[width=0.9\columnwidth]{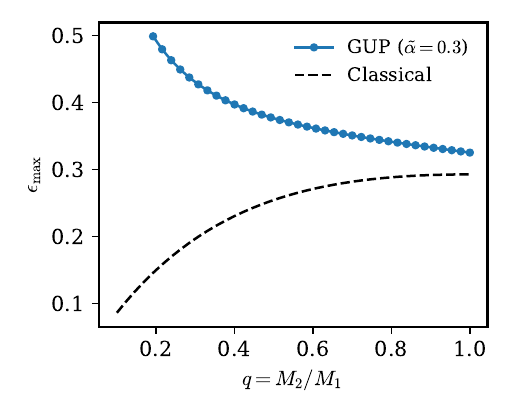}
\caption{
Maximum allowed gravitational-wave efficiency $\epsilon_{\max}$ as a function of the mass ratio $q$, obtained from the condition $\Delta\tilde S = 0$. The solid curve represents the GUP-corrected bound, while the dashed curve corresponds to the classical area-theorem limit. Classical gravity permits larger efficiencies for symmetric mergers, whereas quantum corrections relax the bound for asymmetric systems and impose a more restrictive limit as $q \to 1$.
}
\end{figure}

The maximum allowed efficiency derived from this condition is shown in Figure~2. In the classical case, $\epsilon_{\max}$ increases with the mass ratio and saturates for nearly equal-mass mergers, consistent with expectations from the black hole area theorem. Symmetric mergers generate a larger final horizon and can therefore accommodate greater gravitational-wave emission. When GUP corrections are included, however, the behavior changes qualitatively. The quantum-corrected bound allows higher efficiencies for asymmetric mergers, while imposing a more restrictive upper limit for symmetric systems. This reversal arises from the logarithmic entropy correction, which enhances the entropy contribution of highly unequal mass configurations but becomes less significant as the mass ratio approaches unity. As a result, quantum gravity effects preferentially modify the entropy balance of asymmetric mergers.

\begin{figure}
\includegraphics[width=0.9\columnwidth]{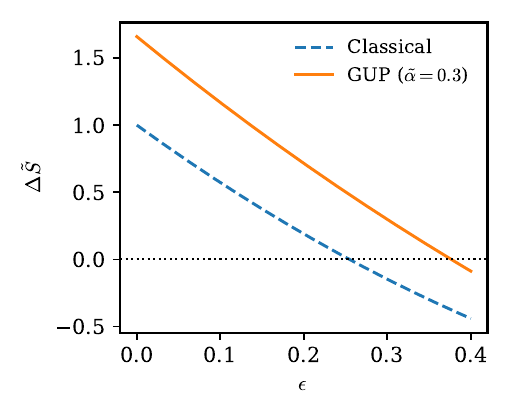}
\caption{
Comparison of the classical and GUP-corrected entropy changes as a function of gravitational-wave efficiency for a fixed mass ratio $q = 0.5$. The GUP correction shifts the entropy curve upward, delaying the onset of entropy violation and extending the range of allowed efficiencies. The horizontal line marks the thermodynamic censorship threshold $\Delta\tilde S = 0$.
}
\end{figure}

Figure~3 isolates this effect by comparing the classical and GUP-corrected entropy changes for a fixed mass ratio. The GUP correction shifts the entropy curve upward, delaying the onset of entropy violation and extending the allowed range of gravitational-wave efficiencies. The approximate parallelism of the two curves indicates that quantum effects act as a perturbative stabilization rather than a qualitative modification of the merger dynamics. This behavior supports the internal consistency of the approach and suggests that the corrections remain under theoretical control.

Overall, our results demonstrate that black hole merger thermodynamics provides a simple yet sensitive probe of quantum gravity corrections. While the predicted bounds do not directly constrain observed gravitational-wave events at present, they establish a consistent theoretical framework in which classical area theorems are smoothly deformed by quantum effects. Extensions of this analysis to spinning black holes, alternative quantum gravity models, or phenomenological parameter constraints may provide further insight into the role of horizon thermodynamics in strong-field gravity.

\section{Conclusions}
In this work, we have investigated the entropy balance of binary black hole mergers by incorporating gravitational-wave energy loss and quantum corrections motivated by the Generalized Uncertainty Principle. By analyzing the transition from the initial binary system to the final merger remnant, we derived thermodynamic bounds on the allowed gravitational-wave efficiency as a function of the mass ratio.

We have shown that, while the classical area theorem is recovered in the appropriate limit, quantum corrections introduce a modified entropy bound that alters the thermodynamic censorship condition. In particular, the logarithmic GUP correction enhances the entropy budget of asymmetric mergers while imposing a more restrictive constraint on nearly equal-mass systems. This behavior highlights the nontrivial interplay between horizon geometry and quantum entropy contributions in black hole merger processes.

Our results indicate that quantum gravity effects can be consistently incorporated into black hole merger thermodynamics without violating the generalized second law. Although the predicted corrections are small and do not directly constrain current gravitational-wave observations, they provide a theoretically controlled framework for exploring deviations from classical gravitational dynamics in strong-field regimes.

\subsection{Application to LIGO--Virgo--KAGRA Events}

To assess the physical relevance of the GUP-modified bounds, we compare our theoretical results with observed compact binary mergers reported in the GWTC catalogs by the LIGO--Virgo--KAGRA collaborations \citep{Abbott2016,Abbott2019}. 

\begin{table}
\centering
\caption{Representative compact binary mergers illustrating the range of mass ratios considered in this work. Mass values are inferred from LIGO--Virgo observations.}
\begin{tabular}{lccc}
\hline
Event & $M_1\ (M_\odot)$ & $M_2\ (M_\odot)$ & $q = M_2/M_1$ \\
\hline
GW150914 & $\sim 36$ & $\sim 29$ & $\sim 0.8$ \\
GW190814 & $\sim 23$ & $\sim 2.6$ & $\sim 0.11$ \\
\hline
\end{tabular}
\end{table}

For a representative near-equal-mass system such as GW150914    ($q \approx 0.86$), the classical area theorem allows a relatively large fraction of the total mass-energy to be emitted in gravitational waves. Within our framework, we find that as $q \to 1$, the inclusion of logarithmic GUP corrections leads to a more restrictive thermodynamic censorship boundary compared to the classical case, thereby reducing the maximum allowed efficiency.

In contrast, highly asymmetric systems such as GW190814 ($q \approx 0.11$) \citep{Abbott2016} exhibit a different behavior. In this regime, the GUP-induced logarithmic correction effectively increases the total entropy budget, which relaxes the corresponding upper bound on gravitational-wave emission relative to the classical limit. Current LVK observations typically report radiated efficiencies $\epsilon \sim 0.05$, well within both classical and quantum bounds. However, the growing deviation between $\epsilon_{\max}^{\mathrm{cl}}$ and $\epsilon_{\max}^{\mathrm{GUP}}$ in the near-equal-mass regime suggests that symmetric mergers provide the most promising setting for probing potential quantum gravity effects. Future high-sensitivity observations, including those anticipated from next-generation detectors such as the Einstein Telescope, may offer improved constraints on such quantum-corrected entropy bounds.

Future extensions of this work may include rotating black holes, alternative quantum gravity-inspired entropy corrections, or phenomenological studies aimed at connecting thermodynamic bounds with numerical-relativity results. Such developments could further clarify the role of horizon thermodynamics as a bridge between classical gravity and quantum gravity effects.

\section{ACKNOWLEDGEMENTS}
The author expresses sincere gratitude to Dr. Antony S (Mahatma Gandhi University) and Dr. R. Tharanadh (Aquinas College, Edakochi) for their valuable support.
\bibliographystyle{abbrv}
\bibliography{References}
\end{document}